# Mapping Academic Institutions According to Their Journal Publication Profile: Spanish Universities as a Case Study

**J. A. García, Rosa Rodríguez-Sánchez, J. Fdez-Valdivia,**

Departamento de Ciencias de la Computación e I. A., Universidad de Granada, 18071 Granada, Spain.

Tel: +34958240592; fax: +34958243317;

Email addresses: jags@decsai.ugr.es; rosa@decsai.ugr.es; jfv@decsai.ugr.es

**N. Robinson-García**[*]

EC3: Evaluación de la Ciencia y la Comunicación Científica, Departamento de Biblioteconomía y Documentación, Universidad de Granada, 18071 Granada, Spain.

Tel: +34958240920; Email address: elrobin@ugr.es

**D. Torres-Salinas**

EC3: Evaluación de la Ciencia y la Comunicación Científica, Centro de Investigación Médica Aplicada, Universidad de Navarra, 31008, Pamplona, Navarra, Spain.

Email address: torressalinas@gmail.com

**Abstract**

We introduce a novel methodology for mapping academic institutions based on their journal publication profiles. We consider that journals in which researchers from academic institutions publish their works can be considered as useful identifiers for representing the relationships between these institutions and establishing comparisons. But, when academic journals are used for research output representation, distinctions must be introduced between them, based on their value as institution descriptors. This leads us to the use of journal weights attached to the institution identifiers. Since a journal in which researchers from a large proportion of institutions published their papers may be a bad indicator of similarity between two academic institutions, it seems reasonable to weight it in accordance with how frequently researchers from different institutions published their papers in this journal. Cluster analysis can then be applied in order to group the academic institutions, and dendrograms can be provided to illustrate groups of institutions following agglomerative hierarchical clustering. In order to test this methodology, we use a sample of Spanish universities as a case study. We first map the study sample according to institutions' overall research output, and then we use it just for two scientific fields (Information and Communication Technologies, as well as Medicine and Pharmacology) as a means to demonstrate how our methodology cannot only be applied for analyzing institutions as a whole, but also in different disciplinary contexts.

*To whom all correspondence should be addressed.

---

## 1. Introduction

Over the last decade a great deal of interest has been focused on scientific mapping and visualization. Although first conceived as tools for displaying the structure and dynamics of research activity, they have now been fully integrated into research evaluation (Noyons, Moed & Luwel, 1999) and combine structural and performance information that enables them as easy-to-read tools for research policy makers (Torres-Salinas, 2009). According to Klavans and Boyack (2009) a map of science can be defined as a set of elements and the existing relationships between them, considering as an element any unit of representation of science such as scientific fields, publications, or researchers. They are characterized by visualizing these elements, commonly represented in a two or three-dimensional space, and by matching pairs of elements according to their common characteristics. Science maps, also known as *Atlas of Science*, are commonly visualized as node-edge diagrams similar to those used in network science and they aim at analyzing the structure of science based mainly in research publications. First attempts to mapping science by applying bibliometric techniques can be traced to Henry Small and his colleagues (Griffith, Small, Stonehill & Dey, 1974; Small, 1999; Small & Garfield, 1985). These techniques vary from each other depending on the methodological choices and on the unit of analysis used.

Although first efforts were made on generating maps based on scientific papers, journals have also been used as a basic unit for mapping science for some 35 years, starting with the pioneering map by Narin, Carpenter and Berlt (1972). These maps are normally generated in two steps. Firstly, a clustering method is used for dividing journals into a number of clusters. The decision made on how these clusters are formed will determine the validity of the whole process as it will define the criteria followed for considering the elements as similar or dissimilar (Gmür, 2003). Secondly, a visualization algorithm is developed in order to generate a layout of the clusters previously formed. In a different approach, Moya-Anegón et al. (2004; 2007) introduced discipline-based maps using the Thomson Reuters subject categories system aiming at a rather ambitious goal such as representing the world's research output. Also Leydesdorff & Rafols (2009) use the Thomson Reuters subject categories for representing science in order to analyze the structure of the Science Citation Index database. Despite technological limitations at first, the emergence since the mid 1990s of new visualization tools and the availability of large amounts of data on scientific publications made possible a further development of this type of maps (Noyons, 2004). Regarding mapping institutions or universities, main efforts have been focused using research collaboration as a means for establishing networks between them (Leydesdorff & Persson, 2010; Rorissa & Yuan, 2012) or web links (Ortega, Aguillo, Cothey & Scharnhorst, 2008), but other than that no other technique has been used. This kind of techniques allow readers to rapidly learn over scientific, geographical, or social connections between different institutions, emphasizing relations that may be crucial on determinant and controversial topics such as the merging of universities (Moed, Moya-Anegón, López-Illescas & Visser, 2011), monitor collaborations and research changes over time (Rafols, Porter & Leydesdorff, 2010) or by extent, any other matter regarding research policy and management at an institutional level (Noyons, 2004).

Taking into account this background, in this paper we propose a novel methodology for representing universities according to their journal publication profile in an attempt to visually synthesize the complex relationships these institutions have with each other. We hypothesize that academic institutions which publish their research output in the same scientific journals should not only have similar research interests but also similar impact, and therefore, should have similar profiles. These last years have seen a great interest on developing measures and thresholds for monitoring and benchmarking universities. The great impact international rankings have had, has not only influenced the Higher Education scenario (Hazelkorn, 2011), but has also risen many questions and critical voices over the methodologies employed when analyzing academic institutions' research output (van Raan, 2005; Torres-Salinas, Moreno-Torres, Delgado-López-Cózar & Herrera, 2011). Universities are subject to numerous influences which differentiate them from other units of analysis such as journals or words. Not only pure research interests drive their relations: geographical and social context among other variables must also be taken into account (Gómez, Bordons, Fernández & Morillo, 2010). In this sense, the application of scientific mapping techniques may be the answer for understanding and reflecting such influences.

This study is structured as follows. In Section 2 we present the proposed methodology for mapping academic institutions. Section 3 describes the sample of 56 Spanish universities used as a case study and tests this novel methodology, applying it over the total scientific output and also focusing only on two areas (Information and Communication Technologies, as well as Medicine and Pharmacology). Section 4 concludes with a discussion over the obtained results.

## 2. Data and methods

The basic idea of the proposed approach is as follows. For each academic institution, we record the scientific journals in which researchers at this institution published their papers during a period of time. No distinction is made between co-authored papers and papers published in a same journal by two different institutions, as we aim at relating universities not just according to their disciplinary focus but also to other external aspects that may influence their similarities such as collaboration or geographical proximity. With the list of scientific journals we construct a journal-by-institution matrix where a given row contains the weights of the corresponding journal across the academic institutions. Here we use the inverse frequency approach (Salton & Buckley, 1988) for generating journal weights, since a journal in which researchers from a large proportion of institutions published their papers should normally be a bad indicator of similarity between two academic institutions. Following a document-document similarity approach (Ahlgren & Colliander, 2009), the behavior of the institution-institution similarity can then be inferred under two types of similarities: first-order and second-order. First-order similarities are obtained by measuring the similarity between columns in a journal-by-institution matrix. However, one may go one step further and obtain them by measuring the similarity between columns in this first-order institution-by-institution similarity matrix. This operation yields a new institution-by-institution matrix, populated with second-order similarities.

In the first-order approach, one focuses on the direct similarity between two academic institutions. The second-order approach determines that, for instance, two universities are similar by detecting that there are other academic institutions such that the two universities are both similar to each of these other institutions. Cluster analysis can then be applied to group the academic institutions in a given set, using second-order institution-institution dissimilarity values. For the cluster analysis here we follow the complete linkage method (Everitt et al., 2001).

*Institution-institution similarities*

Let $U = \{u_i\}$ be a given set of academic institutions under consideration. Here we suggest that the relationships between research output of institutions in $U$ could be represented based on a comparison of academic journals in which researchers from the institutions in $U$ published their manuscripts.

Let $J = \{j_m\}$ be the set of academic journals in which researchers from the institutions in $U$ published their manuscripts during the study time period. Also, let $J_{u_i}$ be the research output of academic institution $u_i$.

With the set of academic journals $J = \{j_m\}$ we construct a journal-by-institution matrix $W = \{w_{m,i}\}$ where a given row contains the weights of the corresponding journal across the academic institutions, in particular, $w_{m,i}$ denotes the weight of journal $j_i$ for representing research output of institution $u_i$.

Following Salton and Buckley (1988), a formal representation of the research output of institution $u_i$ can be obtained by including in $J_{u_i}$ all possible academic journals in J and adding journal weight assignments to provide distinctions among the journals.

Thus if $w_{m,i}$ denotes the weight of journal $j_m$ for representing the research output of institution $u_i$, and a number of M academic journals are available for research output representation, the journal vector for institution $u_i$ can be written as follows:

$$J_{u_i} = \left(j_1, w_{1,i}; j_2, w_{2,i}; \cdots ; j_M, w_{M,i}\right)$$

$$(1)$$

In the following, the basic assumption is that $w_{m,i}$ is equal to 0 when journal $j_m$ is not assigned to institution $u_i$, since researchers of $u_i$ have not published in $j_m$. In order to provide a greater degree of discrimination among journals assigned for research output representation, we also assume that journal weights in decreasing journal importance order could be assigned. Hence, the journal weights $w_{m,i}$ could be allowed to vary continuously between 0 and a maximum allowed value, with the higher weight assignments (near the maximum allowed value) being used for the most important journals regarding research output identification, whereas lower weights near 0 would characterize the less important journals for identification.

Given the journal vector representations in Equation (1), an institution-institution similarity value (that is, an indicator of similarity between two academic institutions $u_i$ and $u_j$ in U) may be obtained by comparing the corresponding journal vectors using the vector product formula. But, the individual journal weights should depend to some extent on the weights of other journals in the same vector. To this aim, it is useful to use normalized journal weight assignments. Using a length normalized journal-weighting system, the institution-institution similarity value reduces to the cosine measure (Baeza-Yates & Ribeiro- Neto, 1999) which gives the cosine of the angle between the two vectors which represent the academic institutions $u_i$ and $u_j$ :

$$B(u_i, u_j) = \frac{\sum_m w_{m,i} \times w_{m,j}}{\sqrt{\sum_m (w_{m,i})^2} \sqrt{\sum_m (w_{m,j})^2}}$$

(2)

where $w_{m,i}$ $(w_{m,j})$ is the weight of journal $j_m$ for research output of institution $u_i$ $(u_j)$; and sums are over all journals in the set $J = \{j_m\}$.

Of course, this is a first-order approach for measuring institution-institution similarities, but the behavior of the institution-institution similarity can be inferred under two types of similarities, first-order and second-order. First-order similarities were obtained in Equation (2) by measuring the similarity between columns in a journal-by-institution matrix $\{w_{mi}\}$, where $w_{mi}$ denotes the weight of journal $j_m$ for institution $u_i$; an operation that yields an institution-by-institution similarity matrix. However, one may go one step further and obtain the similarities by measuring the similarity between columns in this first-order institution-by-institution similarity matrix. This operation yields a new institution-by-institution similarity matrix, populated with second-order similarities. Ahlgren and Colliander (2012) observed good performance of the second-order strategy for measuring similarities in a scientometric context.

From Equation (2), a second order similarity matrix can be defined as follows (Ahlgren & Colliander, 2009):

$$S(u_i, u_j) = \frac{\sum_k B(u_k, u_i) \times B(u_k, u_j)}{\sqrt{\sum_k (B(u_k, u_i))^2} \sqrt{\sum_k (B(u_k, u_j))^2}}$$

(3)

where sums are over all academic institutions in the set *U*.

In designing an automatic institution clustering system, two main questions must be answered. First, what appropriate research output units are to be included in the institution representations? Second, is the determination of the journal weights capable of distinguishing the important journals from those less crucial for research output identification?

Concerning the first question, that is, the choice of research output units, various possibilities may be considered. In this paper, academic journals alone were used for research output representation, given the availability of large amounts of data on scientific publications. However, sets of journals cannot provide complete identifications of research-output. But the judicious use of academic journals for institution representation is preferable when incorporating more complex entities, since the following problems would appear when producing complex identifiers (Salton & Buckley, 1988): (i) Few new identifiers are likely to become available when stringent conditions are used for the construction of complex identifiers; and (ii) many marginal institution identifiers that do not prove useful are obtained when the construction criteria for the complex entities are relaxed. Since the construction and identification of complex institution representations can be inordinately difficult, publication in academic journals was used for research output identification. In order to do so, distinctions must be introduced between individual journals, based on their value as institution descriptors. This leads to the use of journal weights attached to the institution identifiers.

In the next section we consider the generation of effective journal weighting factors.

*Journal weighting system*

A journal-weighting system should increase the effectiveness of institution descriptors. In particular, journals in which researchers from an individual institution frequently published their works appear to be useful as institution identifiers. This suggests that a journal frequency factor can be used as part of the journal-weighting system measuring the frequency of publication in academic journals for a particular institution: $freq_{mi}$ which denotes the number of papers published in journal $j_m$ by researchers at the university $u_i$ during the study time period.

But journal frequency factors alone cannot ensure acceptable institution representation. Specifically, if highly frequent journals are not concentrated in a few particular institutions, but they are prevalent in the whole set $U$, all academic institutions tend to be represented by these same high frequency-journals and it affects the representation precision. Hence a new set-dependent factor must be introduced that favors journals concentrated in a few institutions of the given set $U$. The well-known inverse frequency factor (Salton & Buckley, 1988) can be used to perform this function as follows.

Since a journal in which researchers from a large proportion of institutions published their papers should normally be a bad indicator of similarity between two academic institutions, it is reasonable to weight a journal $j_m$ in accordance with how frequently researchers from different institutions in $U$ published their papers in this journal, for example, by using

$$\log\left(\frac{N}{n_m}\right)$$

(4)

with $N$ being the number of academic institutions in the set $U = \{u_i\}$; and $n_m$ being the number of institutions at which researchers published their work in academic journal $j_m$.

To sum up, the best journals for research-output description are those able to distinguish certain individual institutions from the rest in the given set $U$. This implies that the best journals $j_m$ for representing research output of institution $u_i$ should have high journal frequencies, $freq_{mi}$, but low overall frequencies across institutions in $U$. Following the approach given by Salton and Buckley (1988) and Ahlgren and Colliander (2009), a reasonable measure of journal importance may then be obtained by using the product of the journal frequency and the inverse frequency factor. Let $j_m$ be the m-th considered academic journal in $J$. We now define the weight of journal $j_m$ for representing research output of institution $u_i$ as:

$$w_{m,i} = freq_{mi} \times \log\left(\frac{N}{n_m}\right)$$

(5)

where $freq_{mi}$ is the number of papers published in journal $j_m$ by researchers at the university $u_i$ during the time period under consideration; and the inverse frequency factor $\log\left(\frac{N}{n_m}\right)$ varies inversely with the number of institutions at which researchers published their work in the same journal $j_m$.

*Assigning a set of academic institutions into groups*

Cluster analysis can then be applied in order to group the academic institutions in *U*. To this aim, similarity values obtained by Equation (3) are firstly converted to corresponding dissimilarity values by subtracting a given similarity value from 1. For the cluster analysis, we follow the complete linkage method (Everitt et al., 2001). In cluster analysis, complete linkage or furthest neighbor is a method for calculating distances between clusters in agglomerative hierarchical clustering. In complete linkage, the distance between two clusters is computed as the maximum distance between a pair of objects, one in one cluster, and one in the other, (Everitt et al., 2001). Thus, the distance between two clusters of academic institutions, $C1$ and $C2$, is defined as the maximum dissimilarity between two institutions *u* and *v*, where $u \in C1$ and $v \in C2$:

$$D(C1, C2) = \max_{u \in C1; v \in C2}\big(d(u,v)\big)$$

For example, complete linkage clustering, based on the generated dissimilarity matrices, can be performed following MathWorks (2012).

In agglomerative hierarchical clustering, the clusters are initially the single-member clusters. At each stage the academic institutions or groups of institutions that are closest according to the linkage criterion are joined to form a new, larger cluster. At the last stage, a single group consisting of all academic institutions is formed. This avoids the problem of determining the number of clusters which is often ambiguous, with interpretations depending on the shape and scale of the distribution of points in a data set and the desired clustering resolution of the user. The components at each iterative step are always a subset of other structures. Hence, the subsets can be represented using a tree diagram, or dendrogram. Horizontal slices of the tree at a given level indicate the clusters that exist above and below a value of the weight. Maps of academic institutions are node-edge diagrams, locating each institution in a two or three-dimensional space and with the explicit linking of pairs of institutions by virtue of the relationships between them, i.e., institution-institution similarities. In addition, dendrograms can be provided to illustrate the clustering of institutions or groups of institutions following agglomerative hierarchical clustering, (MathWorks, 2012). Table 1 summarizes the methodological approach for construction of maps of academic institutions and the corresponding dendrograms.

TABLE 1. Sum of the proposed methodology for mapping universities according to their journal publication profile

| *Algorithm 1 Methodological procedure* |
| --- |
| 1. Obtain list of journals on which each institution has published for the study time period |
| 2. Apply weights to journals for each institution according to Equation (5). |
| 3. Construct a journal-by-institution matrix. |
| 4. Extract values from an institution-institution matrix derived from Equation (1). |
| 5. Apply a second-order approach to emphasize similarities among institutions. |
| 6. Perform a complete linkage clustering method in order to set the institutions groups according to their journal publication profile. |
| 7. Construct a dendrogram with all university groups |
| 8. Map the universities network according to their similarity |

*Data source and processing*

Considering that the aim was to visualize the relationships between universities based on their scientific production, the Thomson-Reuters Web of Science database was selected as data source. This decision is based on the great regard this database has for research policy makers, as it is considered to store the most relevant scientific literature in the world. Then, a set of academic institutions selected according to their research output and a study time period were chosen. We manually performed a search query for each university in order to download their research output data. For this, we used the 'Address' filter taking into account all possible names for each institution. Then, we downloaded all records assigned to each institution. We only considered as scientific publications those belonging to journals indexed in one of the Thomson-Reuters Journal Citation Reports (hereafter JCR). These lists of journals are divided per subject categories and contain several bibliometric indicators. One of them is the Impact Factor, which is used as a ranking indicator for ordering journals according to their impact in scientific literature. The editions of the JCR for the study time period were downloaded in September 2011. Also, we calculated the percentage of papers indexed in fist quartile journals (hereafter Q1 journals). Despite not being necessary for reproducing the suggested methodology, we considered that introducing a color range depending on the percentage of publications in Q1 journals would enrich the maps and ease our discussion over the results when demonstrating how it does not only group universities according to their disciplinary focus but also to their capability on publishing in top journals. This should not be interpreted as assuming that certain universities publish papers of higher impact than others (García, et al, 2012a) but as a competitive advantage of its researchers in terms of visibility.

**3. Case study: Map of Spanish universities based on institution-institution similarities**

*3.1. Global map of Spanish universities*

As a means of validating and applying the proposed methodology for mapping universities (see Table 1), we selected a set of Spanish universities with at least 50 citable documents (articles, reviews, notes and letters) published in JCR Journals, resulting in 56 universities (see Table 2), and downloaded their production for the 2008-2010 time period. The timeframe chosen aims at portraying as accurately as possible the current Spanish higher education landscape regarding its research performance. For each university we retrieved all scientific journals in which researchers from each institution published their papers during the study time period. We then used the cosine measure to compute a first-order and second-order similarity between universities. The map of Spanish universities will be a node-edge diagram, locating each university in a two-dimensional space and with the explicit linking of pairs of universities by virtue of the relationships between them, i.e., university-university similarity values. For this, the software program Pajek (Networks/Pajek, 2011) was used and universities' positioning was determined in accord to the Kamada-Kawai algorithm (Kamada-Kawai, 1998), which is commonly used in this kind of representations. Next, we used the complete linkage method for clustering the 56 Spanish universities using second-order dissimilarities.

TABLE 2. Set of Spanish universities used as sample for mapping institutions according its scientific research output during de 2008-2010 time period

| University | NDOCS | %Q1 | University | NDOCS | %Q1 | University | NDOCS | %Q1 |
|---|---|---|---|---|---|---|---|---|
| BARCELONA | 11168 | 56% | ALICANTE | 2349 | 50% | LLEIDA | 1124 | 51% |
| AUTÓNOMA DE BARCELONA | 8428 | 56% | CÓRDOBA | 2334 | 57% | ALMERIA | 1085 | 46% |
| COMPLUTENSE MADRID | 7629 | 51% | ROVIRA I VIRGILI | 2302 | 55% | PUBLICA DE NAVARRA | 1016 | 44% |
| VALENCIA | 6764 | 54% | VALLADOLID | 2187 | 43% | PALMAS (LAS) | 1016 | 43% |
| AUTÓNOMA DE MADRID | 6386 | 56% | LAGUNA, LA | 2176 | 52% | UNED | 929 | 41% |
| GRANADA | 5380 | 49% | MALAGA | 2076 | 48% | LEON | 917 | 48% |
| POLITÉCNICA DE CATALUÑA | 4992 | 49% | POMPEU FABRA | 1972 | 59% | POLITÉCNICA CARTAGENA | 908 | 46% |
| PAIS VASCO | 4827 | 52% | CANTABRIA | 1826 | 51% | HUELVA | 748 | 52% |
| ZARAGOZA | 4487 | 53% | EXTREMADURA | 1816 | 49% | PABLO OLAVIDE | 656 | 51% |
| SEVILLA | 4484 | 50% | ALCALA DE HENARES | 1809 | 46% | BURGOS | 478 | 52% |
| POLITECNICA DE VALENCIA | 4445 | 49% | CARLOS III | 1805 | 43% | RIOJA (LA) | 446 | 50% |
| SANTIAGO DE COMPOSTELA | 4400 | 50% | ISLAS BALEARES | 1565 | 56% | RAMON LLUL | 366 | 38% |
| POLITÉCNICA DE MADRID | 4065 | 43% | GIRONA | 1520 | 53% | EUROPEA DE MADRID | 190 | 45% |
| OVIEDO | 3232 | 49% | MIGUEL HERNANDEZ | 1519 | 48% | CARDENAL HERRERA-CEU | 189 | 34% |
| VIGO | 2983 | 49% | REY JUAN CARLOS | 1512 | 49% | SAN PABLO CEU | 171 | 49% |
| CASTILLA LA MANCHA | 2829 | 50% | CORUÑA, A | 1439 | 47% | PONTIFICIA COMILLAS | 144 | 45% |
| MURCIA | 2663 | 45% | JAEN | 1355 | 43% | MONDRAGON | 80 | 39% |
| SALAMANCA | 2510 | 48% | CADIZ | 1261 | 48% | DEUSTO | 55 | 22% |
| NAVARRA | 2469 | 47% | JAUME I | 1225 | 54% | | | |

**Indicators:**

**NDOCS**: Number of citable documents (article, review, note or letters) indexed in JCR Journals (Thomson-Reuters)

Here we have used the cosine measure to compute the first-order and second-order similarity between universities as given above (see Equations (2) and (3)). The second-order similarity matrix S contains many cells with very low similarities. From a computational point of view, it is problematic to keep all such similarities in the matrix. Moreover, to take them into account in the computations might have a negative impact on the visualization quality. We handled this problem by establishing minimum similarity values (e.g., 0.6 in Fig. 1).

Figure 1 shows the resulting map for Spanish universities. Four distinct groups of universities can be inferred according to similarities in their research profile. On the first hand we have a group formed by the five universities which could be considered as the most important ones (Barcelona, Autónoma de Madrid, Autónoma de Barcelona, Valencia and Complutense Madrid) as these occupy the highest positions (for Spanish universities) in well-known international rankings such as the Shanghai Ranking (Shanghai Jiao Tong University, 2011) or the Performance Ranking of Scientific Papers for World Universities (Higher Education Evaluation & Accreditation Council of Taiwan, 2011). These universities are the ones with the highest production and more links with the rest of universities which seem to surround them. The high number of links may suggest that they are not just highly productive universities, but also generalist universities covering different disciplines. It is also noticeable that, except Valencia, all universities belong either to Madrid or Barcelona, the two main cities in Spain. They are similar universities not only in their disciplinary orientation, but also in their size and scientific impact according to its percentage of documents in Q1 journals. The second group (Granada, Santiago, Zaragoza, País Vasco, Sevilla) would be formed by a set of universities also generalist and surrounded by a dense network but of a smaller size. Funnily enough these universities usually occupy positions between 400-500 in the Shanghai Ranking; dropping out some years and appearing others, which also reinforces their similarity. However, some distinctions can be made when relating their Q1 production and their positions in the Shanghai Ranking; while Granada appears in all editions of the ranking, the others drop in some editions, maybe related to the proportion of Q1 production each university has. In this sense, it seems that this university is somewhere between these two groups.

FIGURE 1. Map of main Spanish universities according to their journal publication profile.

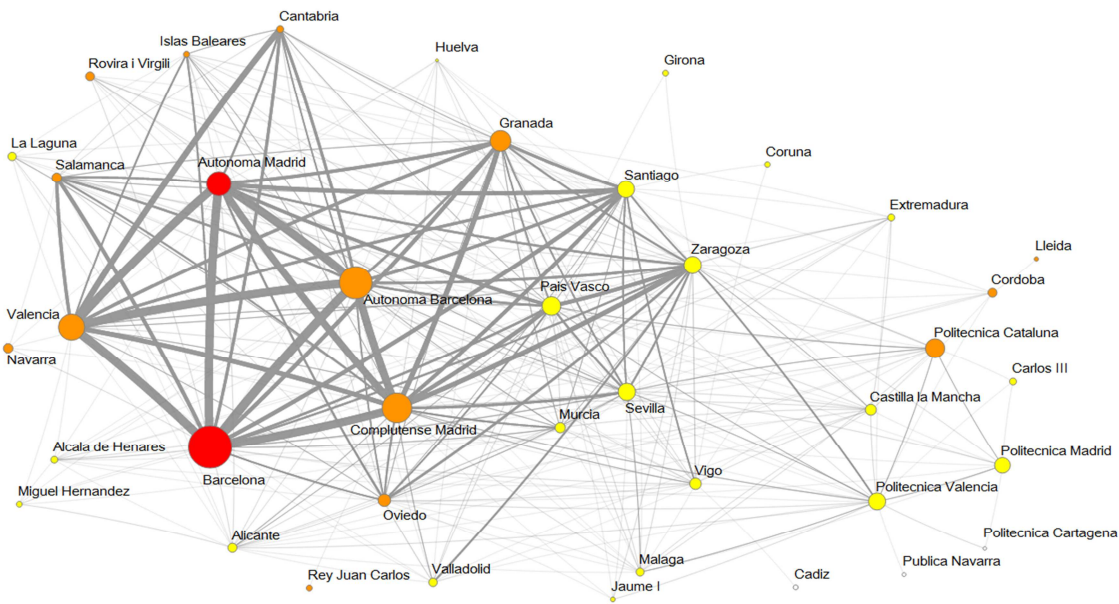

**Map Characteristic:** Lines > minimum similarity value 0.60; maximum similarity value 0.98. **Isolated university nodes** have been removed. From 0.75 line-width is emphasized. **Colors:** ■ >50% production belongs to Q1 journals; ■ 40-50% production belongs to Q1 journals; ■ 30-40% production belongs to Q1 journals; □ <30% production belongs to Q1 journals.

A third group can be distinguished by less productive universies (hence, smaller universities) which have strong links only with those universities belonging to the first group, showing similarities in certain fields of endeavor. These universities are characterized by their size. They seem to reflect the model of bigger universities and therefore their similarities with these universities. Universities belonging to this group would be Cantabria, Islas Baleares or Oviedo for instance. The fourth group is integrated by small universities with weak links to universities belonging to the first or second group. These weak links are  due to a high specialization on certain fields also common to the other universities (Torres-Salinas, Delgado-López-Cózar, Moreno-Torres & Herrera; 2011). An example of this would be Navarra (Medicine and Pharmacy), Rovira i Virgili (Chemistry), or Murcia (Biological Sciences). The last group is mainly formed by the universities named as Polytechnics or Technological (Politécnica de Madrid, Politécnica de Valencia, Politécnica de Cataluña, etc.). Though these universities are linked with the rest of universities, they are also linked between them. The reason for showing such weak links is  due to their high specialization on certain scientific fields belonging to the Engineering and Applied Sciences. In fact, surrounding them we also find other universities that show a tendency towards this "technological" profile, such as Zaragoza (which shares a strong link to Politécnica de Valencia), Carlos III, Pública de Navarra or Castilla La Mancha.

The high minimum values established in Figure 1, seem to eliminate most  reflections of the geographical or regional relations among universities, emphasizing purely research similarities. But we can still trace this kind of relationship between three universities: Santiago de Compostela, Vigo and Coruña. In this case, the interpretation seems to be quite reasonable. The two latter universities were formed in 1990 and 1989 respectively both from campuses belonging to the former university, which is a historical university funded in the fifteenth century.

FIGURE 2. Dendrogram of Spanish universities according to their journal publication profile

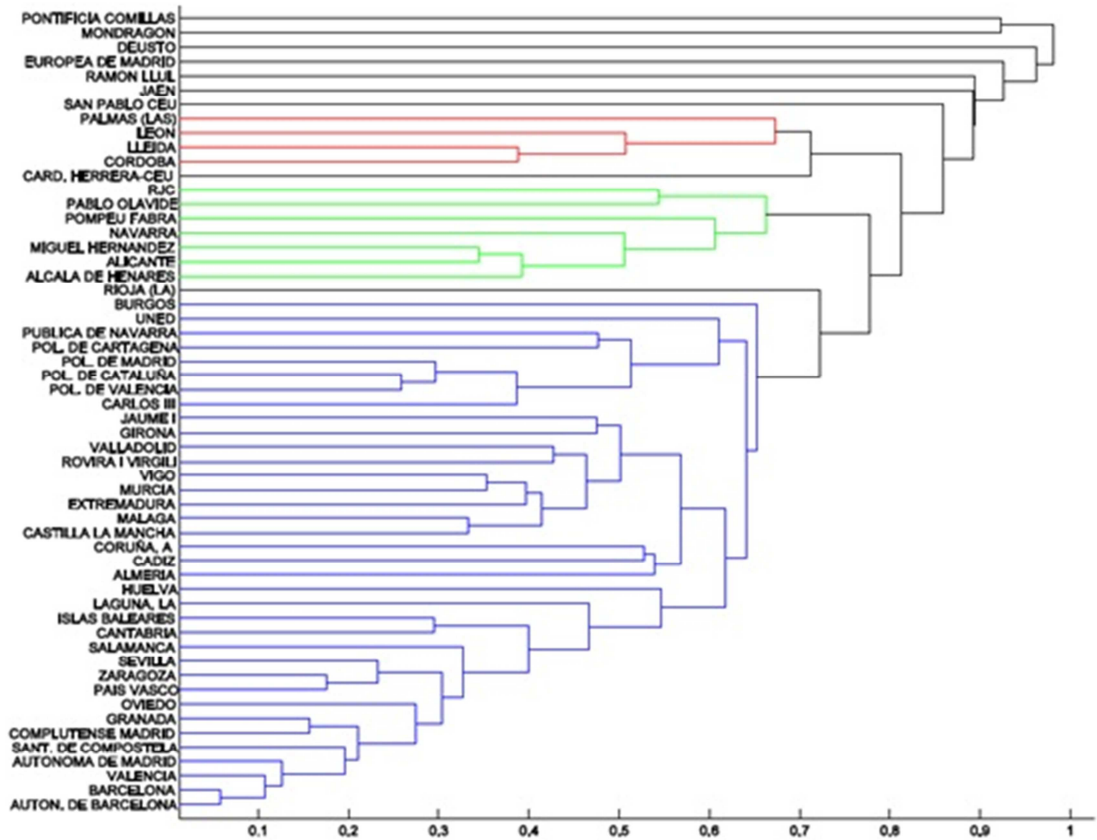

In this map we find that one important university is missing, the University of Pompeu Fabra. This Catalan university has experienced a meteoric growth during the last years. A relatively new university (it was founded in 1990), during the last two years it has appeared in the most renown international rankings: between the 300 and 500 top class universities according to the Shanghai Ranking since 2009 or between the 150 and 200 top universities in the last two years according to The Times World Universities Ranking, for instance. Its absence in Figure 1 suggests that its publication patterns differ from the rest of the Spanish universities, suggesting that probably its journal publication profile may be oriented in such a way that can explain such an outburst. As we indicated before, by using common journals as a means for mapping universities, we not only group them according to their research profile, but also to their research impact (understood as the impact factor of journals in which their output is published). This university serves as a good example of this second characteristic as 59% of its production is published in Q1 journals (see Table 2), that is the highest proportion for the sample used. This way we can see how its absence may not have to do so much by its disciplinary profile but with the journals in which it publishes. Figure 2 shows a dendrogram of Spanish universities or groups of universities following agglomerative hierarchical clustering. From this figure, it follows the rapid grouping of Barcelona, Autónoma de Barcelona, Valencia, and Autónoma de Madrid, which belong to the core of the map of Spanish universities according to their journal publication profile as given in Fig. 1. We have also that Granada and Complutense de Madrid form a very strong grouping. Another relatively natural grouping is formed by Politécnica de Valencia, Politécnica de Cataluña and Politécnica de Madrid, all of them which are universities with a tendency towards the technological profile. From Fig. 2, we have that Sevilla, Zaragoza, and País Vasco belong to another group of universities according to their journal publication profile.

*Specific maps of Spanish universities for the fields of Information and Communication Technologies, as well as Medicine and Pharmacology*

After testing our methodology for the total production of universities, we go a step further and test it for different scientific fields in the belief that in order to have a clear and more precise picture of universities' similarities, it is necessary to deepen on specific fields so that we can understand better their relations. For this, we focus in two different areas: Information and Communication Technologies (hereafter ICT) and Medicine and Pharmacology (hereafter MED). We construct these fields by aggregating thematically the Thomson Reuters subject categories, following the same criteria we did in a previous study[1] (Torres-Salinas, Moreno-Torres, Robinson-García, Delgado-López-Cózar & Herrera; 2011). We use the same set of 56 Spanish universities (Table 2) and the same study time period (2008-2010).

FIGURE 3. Map of Spanish universities according to their journal publication profile in ICT

---

[1] For a better understanding on how these broad scientific fields were formed the reader is referred to http://www.ugr.es/~elrobin/rankingsISI_2011.pdf where we show the correspondence followed between the ISI subject categories and 12 scientific fields including the two used in this study.

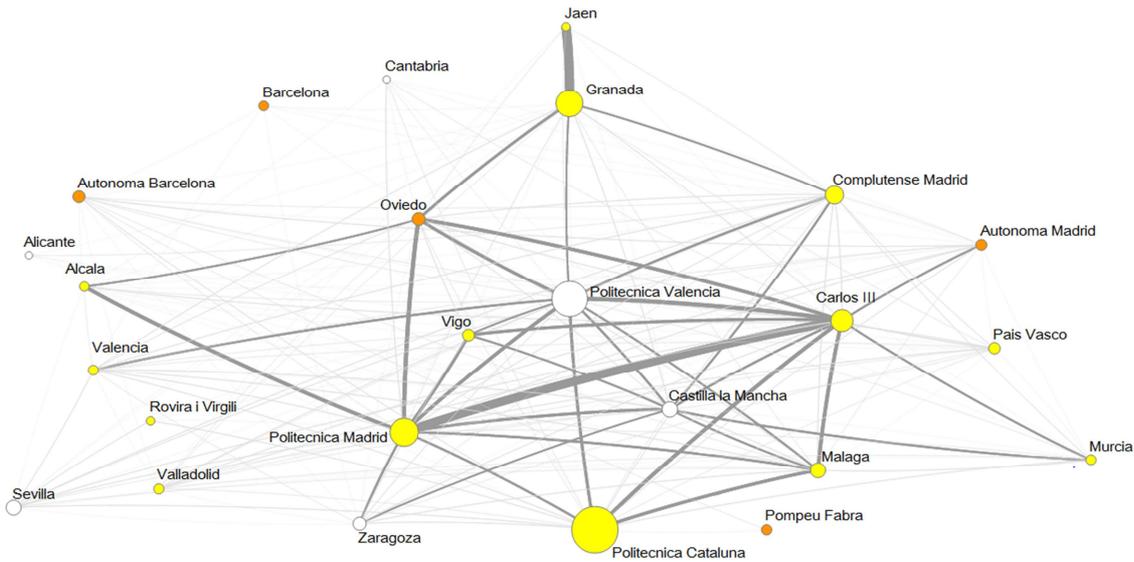

**Map Characteristic:** Lines > minimum similarity value 0.60; maximum similarity value 0.875. **Isolated university nodes** have been removed. From 0.75 line-width is emphasized. **Colors:** 🟥 >50% production belongs to Q1 journals; 🟧 40-50% production belongs to Q1 journals; 🟨 30-40% production belongs to Q1 journals; ⬜ <30% production belongs to Q1 journals.

In Figure 3 we map Spanish Universities according to their journal publication profile in ICT. In this case, disciplines are crucial on shaping universities similarities. We find that Politécnica de Valencia shows a much more diversified profile in this scientific field, occupying a central place in the representation. That is, it is similar to a greater amount of universities, signifying its lesser specialization on certain disciplines. Oviedo, Politécnica de Madrid and Carlos III show greater similarities among them and also, each of them is the core for grouping other universities.

But the most interesting patterns are those followed by Granada and Politécnica de Cataluña. According to their research impact and output, these two universities are the top ones on this scientific field (Torres-Salinas, Delgado-López-Cózar, Moreno-Torres, Herrera; 2011) but they are not the core of the representation as one would have thought. Instead, they seem to follow different patterns than the rest of the universities, suggesting a highly specialized profile in both cases. While Politécnica de Cataluña shows stronger similarities with other universities such as Málaga, Carlos III, Politécnica de Madrid and Politécnica de Valencia; Granada shows a high similarity with Jaén and weaker ones with the rest. The reason for this dissimilarity could lay on a high specialization on different research lines than those followed by the rest of the universities. Also there are geographical and social factors that influence the strong similarity with Jaén among those related with research. As it occurred with Santiago de Compostela, Vigo and Coruña before, Jaén is a relatively new university (it was founded in 1993) which used to be a campus belonging to the University of Granada. This social context may explain their similarity, as there are probably still strong collaboration links between researchers in ICT belonging to both universities.

FIGURE 4. Detail of disciplinary differences in ICT between Granada, Jaén and Politécnica de Cataluña according to the Thomson Reuters subject categories

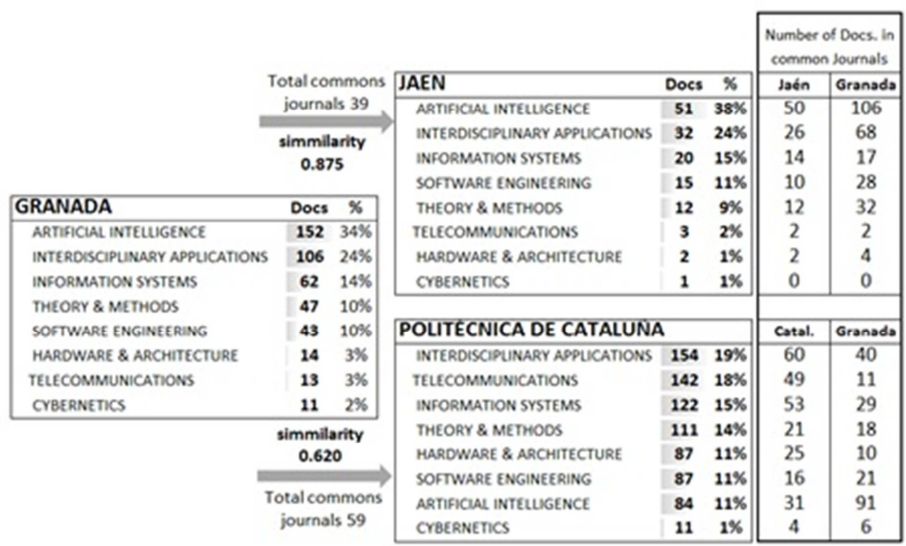

This hypothesis is reinforced by Figure 4 in which we see the distribution of research output according to the Thomson Reuters Subject Categories for three universities: Granada, Jaén and Politécnica de Cataluña. Deepening in categories allows us to observe the similarities between the two former and dissimilarities with the latter. This way we see how high levels of similarity correspond with similar publication profiles; Jaén and Granada's research distribution per categories is very similar and much focused in two main categories (*Artificial intelligence* and *Interdisciplinary applications*) which contain more than half of their total production for both universities. On the other hand, Politécnica de Cataluña shows a more diversified profile never reaching 20% of its production in just one category. It is also interesting to see how the proposed methodology is not influenced by size. Despite having Granada more journals in common with Politécnica de Cataluña, the proportion of publications in the same journals with Jaén is higher, which explains their similar profile.

When focusing in MED a different picture emerges (Figure 5), signifying how necessary becomes a disciplinary approach to universities when establishing research profiles. In this case we find four distinct groups of universities. The main one is composed by Barcelona, Autónoma de Barcelona and Autónoma de Madrid, which have strong similarities among them. They are characterized by their large production and by publishing in Q1 journals (only Autónoma de Barcelona has less than half of its output published in Q1 journals). They are also the most generalist universities in this field of endeavor as they represent the core of the map. Then, we find a second group of universities with high outputs which surround this core (Complutense de Madrid, Navarra, Valencia). In the case of Navarra and comparing with Figure 1, it is plausible to suggest that it is a highly specialized University in MED with a very similar profile to Autónoma de Madrid, Barcelona, Autónoma de Barcelona and Valencia. The third group is formed by universities with weak links with universities belonging to the other two groups, for instance, Alcalá de Henares, Granada or País Vasco.

FIGURE 5. Map of Spanish universities according to their journal publication profile in MED

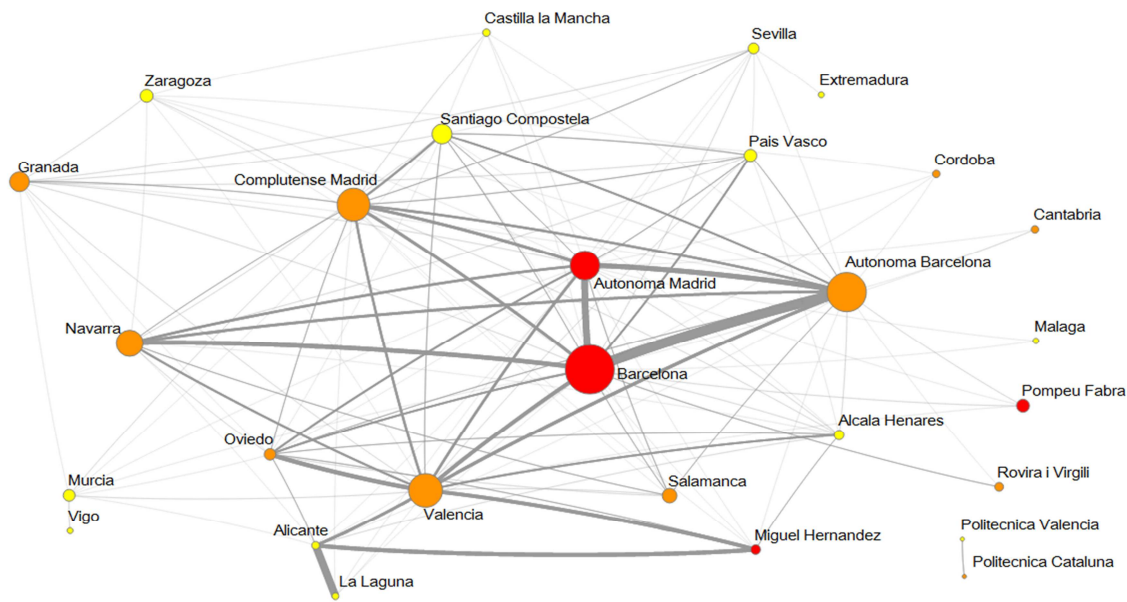

**Map Characteristic:** Lines > minimum similarity value 0.60; maximum similarity value 0.93. **Isolated university nodes** have been removed. From 0.75 line-width is emphasized. **Colors:** ■ >50% production belongs to Q1 journals; ■ 40-50% production belongs to Q1 journals; ■ 30-40% production belongs to Q1 journals; □ <30% production belongs to Q1 journals.

It is worth mentioning a fourth group formed by just two universities and completely separated from the rest. This is the one formed by Politécnica de Valencia and Politécnica de Cataluña. As it can be drawn through all this section, Polytechnics are very similar in their research profile. In this case, this similarity between them on the one hand, and dissimilarity from the rest of the universities on the other, is due to a research interest focused on the *Engineering, Biomedical* Thomson Reuters JCR subject category which would explain why there is no connection with the other universities. In fact, their production in this category represent 30% of their total output in MED, that is, 61 documents published by Politécnica de Cataluña and 66 documents published by Politécnica de Valencia.

In Figure 6 we emphasize as we did with ICT (Figure 4), the capability of the proposed methodology for grouping similar universities and separating dissimilar universities according to their journal publication profile in MED. In this case, we compare the distribution of research output according to the Thomson Reuters Subject Categories of Autónoma de Barcelona with Barcelona and Alcalá de Henares. That is, with its most similar university and a lesser similar one. In the first case, we observe a similarity of 0.930, which stresses how alike the profile of these two universities is in this scientific field. In fact, the eight categories in which they produce more documents are the exact same for both institutions. On the other hand, when comparing Autónoma de Barcelona with Alcalá de Henares we see that, despite publishing an important proportion of their total output in the same four categories, - mainly those related with Neurosciences, - they also present a special focus on different specialties that make them quite different (in the case of Alcalá de Henares for instance, *Ophthalmology*, *Oncology* or *Surgery*). Thereby we can witness once more how the methodology employed groups universities according to their research and publication similarities.

FIGURE 6. Detail of disciplinary differences in MED between Autónoma de Barcelona, Barcelona and Alcalá de Henares according to the Thomson Reuters subject categories

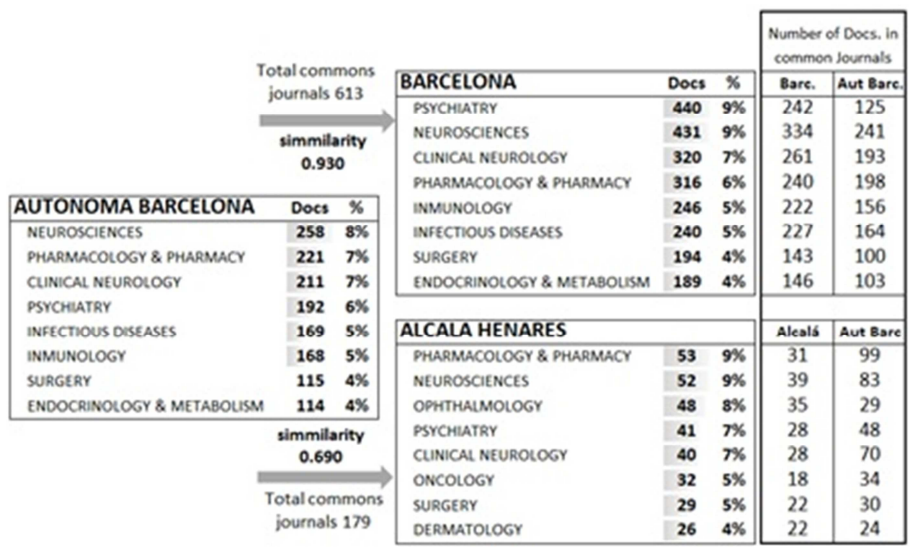

## 4. Discussion and concluding remarks

The present study aims at proposing a novel methodology for mapping academic institutions according to their research profile. Based on the presumption that similar universities should publish in the same scientific journals, we present an algorithm for measuring similarities between universities and their journal publication profile and we represent them in a dendrogram and a network map. In order to test this methodology we set a sample of 56 Spanish universities and a three-year study time period (2008-2010). Then, we apply this methodology in three different scenarios: a representation of universities according to their total output, a representation according to their output in ICT, and a representation according to their output in MED.

This way we first analyze its potential for grouping institutions in a competitive context deeply influenced by table leagues and rankings in which it has repeatedly been noted that only similar institutions can be compared in order to proceed properly when ranking (van der Wende & Westerheijden, 2009). This can be seen in Figure 1 where we observe how the proportion of publications in Q1 journals for universities is similar for each of the previously discussed groups. Although some attempts have been done when classifying universities according to their research performance (Shin, 2009), this approach focuses on mapping universities according to their journal publication profiles, in the belief that this perspective ends with limitations derived from a rigid classification system subjected to a fixed set of criteria. Also, it allows grouping universities taking into account their disciplinary similarities (Lopez-Illescas, Moya-Anegón & Moed, 2011) and their research impact or quality (considering as such publications in Q1 journals). This way we address not only to vertical diversity between universities, which is the one rankings emphasize, but also horizontal diversity.

In this vein go the other two tests presented. When analyzing the methodology in two different scientific fields, we intend to demonstrate how our approach can, not just group similar universities, but also detect similarities between institutions that are centered in the same disciplines and specialties. Also, we have noted that, having a previous knowledge over a determined higher education system over which the procedure is performed, we can also discover geographical, social and/or historical relationships between academic institutions, as we have previously seen in the case of Santiago de Compostela, Vigo and Coruña in Figure 1 or Granada and Jaén in Figure 3.

To validate the results illustrated in Figure 1, a different method with similar results needs to be presented. We used García et al (2012b) where a summary measure of multidimensional prestige of influential fields was introduced to assess the comparative performance of Spanish Universities during the period 2006-2010.

To this aim, a field of study at a given university is considered as having dimension specific prestige when its score based on a given ranking model (e.g., %Q1) exceeds a threshold value. Then, it can be defined which fields at a given university are considered to be prestigious in a multidimensional setting. Thus, a field of study at this university has multidimensional prestige only if it is an influential field with respect to a number of dimensions. Finally, after having identified the multidimensional influential fields at a particular university, their prestige scores are aggregated to a summary measure of multidimensional prestige. The summary measure is not only sensitive to the number of dimensions but also takes into account changes in the ranking scores of influential fields of study at the university.

García et al (2012b) shows the ranking of research output of Spanish universities during the period 2006-2010 (see Table 5). To this aim it was computed the multidimensional prestige of influential fields of study at each institution using a multivariate indicator space. Six variables were used in this analysis: (1) Raw number of citable papers (articles, reviews, notes or letters) published in scientific journals (NDOC); (2) Number of citations received by all citable papers (NCIT); (3) H-Index (H); (4) Ratio of papers published in journals in the top JCR quartile (%Q1); (5) Average number of citations received by all citable papers (ACIT); and (6) Ratio of papers that belong to the top 10% most cited (TOPCIT). The data are available at http://www.rankinguniversidades.es. Fifty-six main universities in Spain are considered in this experiment.

From the results shown in García et al (2012b), the top 8 Spanish universities during the period 2006-2010 were: (1) Barcelona; (2) Autónoma de Barcelona; (3) Autónoma de Madrid; (4) Valencia; (5) Complutense de Madrid; (6) Granada; (7) Santiago de Compostela; and (8) Zaragoza. Also it follows that País Vasco and Sevilla are very similar according to their multidimensional prestige of influential fields. This also happens to two other technological universities: Politécnica de Valencia and Politécnica de Cataluña; which are similar according to their multidimensional prestige (see Table 5 in that paper).

The interesting point is that all these results are congruent with those from the present study (as given by Figure 1 and Figure 2) where we analyze the main Spanish universities according to their journal publication profile.

This type of representation offers a new model for visualizing universities' relationships that can show more clearly than other types of mapping (such as collaboration or web-links maps) the multidimensional similarities and dissimilarities between academic institutions. Likewise, this tool serves as a perfect complement for interpreting universities' performance in rankings as a means for understanding them not as isolated entities, but as interrelated elements of a national higher education system. At a research policy level, this mapping technique may be of use when identifying and selecting universities with similar profiles, as it helps us to identify which universities can be compared and which not, not just at a national level, as has been described through all the paper, but also to compare universities at a transnational or international level. Finally, in the national context it may be of special interest for research policy managers when analyzing potential merging of universities or concentration of research. This last idea goes in consonance with recent developments in Spain regarding its research policy and the 'International Excellence Campus' [Campus de Excelencia Internacional] program which aims at encouraging universities' collaboration.

However, some limitations have also been noted. Using the journal publication approach we find too many links between universities, which makes it difficult to visualize universities under certain levels of similarity, blurring similarities between low performance universities. This limits the analysis when mapping a whole national higher education system as some universities have to inevitably, drop out. In this sense, it also understandable that applying this type of methodologies under a certain threshold is not advisable. Also it would be of interest to introduce other document types (monographs for instance) that could permit a better coverage of certain fields such as social sciences and humanities, and develop methodologies that would adjust to these document types.

**Acknowledgments**

The authors would like to thank the two anonymous reviewers for their constructive suggestions and thoughtful comments. This research was sponsored by the Spanish Board for Science and Technology (MICINN) under grant TIN2010-15157 co-financed with European FEDER funds. Nicolás Robinson-García is currently supported by a FPU grant from the Ministerio de Educación y Ciencia of the Spanish Government.